\documentclass[twocolumn]{aastex631}

\usepackage{amsmath}
\usepackage{xspace}
\usepackage{xcolor}
\usepackage{multirow}

\newcommand{\Vsquared}{V$^2$\xspace}

\newcommand{\hMpc}{Mpc/$h$\xspace}
\newcommand{\hGpc}{Gpc/$h$\xspace}

\shorttitle{Crossing numbers in voids}
\shortauthors{Veyrat et al.}

\begin{document}

\title{A Comparison of Void-Finding Algorithms using Crossing Numbers}

\correspondingauthor{Dahlia Veyrat}
\email{dveyrat@ur.rochester.edu}

\author[0000-0001-8101-2836]{Dahlia Veyrat}
\affiliation{Department of Physics \& Astronomy, University of Rochester, 500 Wilson Blvd., Rochester, NY  14627}

\author[0000-0002-9540-546X]{Kelly A. Douglass}
\affiliation{Department of Physics \& Astronomy, University of Rochester, 500 Wilson Blvd., Rochester, NY  14627}
\email{kellyadouglass@rochester.edu}

\author[0000-0001-5537-4710]{Segev BenZvi}
\affiliation{Department of Physics \& Astronomy, University of Rochester, 500 Wilson Blvd., Rochester, NY  14627}
\email{segev.benzvi@rochester.edu}

\begin{abstract}
  We study how well void-finding algorithms identify cosmic void regions and 
  whether we can quantitatively and qualitatively compare the voids they find 
  with dynamical information from the underlying matter distribution.  Using the 
  ORIGAMI algorithm to determine the number of dimensions along which dark 
  matter particles have undergone shell-crossing (crossing number) in $N$-body 
  simulations from the AbacusSummit simulation suite, we identify dark matter 
  particles that have undergone no shell crossing as belonging to voids.  We 
  then find voids in the corresponding halo distribution using two different 
  void-finding algorithms: VoidFinder and \Vsquared, a ZOBOV-based algorithm.  
  The resulting void catalogs are compared to the distribution of dark matter 
  particles to examine how their crossing numbers depend on void proximity.  
  While both algorithms' voids have a similar distribution of crossing numbers 
  near their centers, we find that beyond 0.25 times the effective void radius, 
  voids found by VoidFinder exhibit a stronger preference for particles with low 
  crossing numbers than those found by \Vsquared.  We examine two possible 
  methods of mitigating this difference in efficacy between the algorithms.  
  While we are able to partially mitigate the ineffectiveness of \Vsquared by 
  using the distance from the void edge as a measure of centrality, we conclude 
  that VoidFinder more reliably identifies dynamically distinct regions of low 
  crossing number.
\end{abstract}

\section{Introduction}

On cosmic scales, the structure of matter in the observable universe exhibits a 
complex, web-like distribution \citep{Bond96}, with distinct filaments 
stretching between dense clusters of galaxies.  In the space between these 
superstructures, galaxy redshift surveys have found vast, relatively empty 
regions containing very few galaxies \citep{Joeveer78, Gregory78, Kirshner81}.  
These cosmic voids, which occupy most of the volume of the universe 
\citep{deLapparent86, Geller89}, are the result of the gravitational instability 
in primordial underdense regions, similar to the formation of clusters from 
gravitational collapse of primordial overdensities \citep{Zeldovich70, 
vandeWeygaert11}.

The emptiness of voids provides a unique environment of gravitational evolution 
for cosmological and astrophysical studies.  Cosmologically, the lack of 
large-scale gravitational collapse causes dynamics within voids to remain in the 
linear regime for a relatively longer time \citep{Goldberg04}.  Their uniqueness 
as cosmic structures has made voids well-suited to studies of the 
Alcock-Paczy{\'n}ski effect \citep[e.g.][]{Lavaux12, Sutter12b, Sutter14b, 
Hamaus16, Mao17a, Nadathur19}, dark energy \citep[e.g.][]{Pisani15, Verza19}, 
baryon acoustic oscillations \citep[e.g.][]{Nadathur19, Zhao20, Zhao22}, and 
weak lensing \citep[e.g.][]{Melchior14, Chantavat17}.  Additionally, the 
uniqueness of voids as an intergalactic environment results in measurably 
different evolution and properties of galaxies within them 
\citep[e.g.][]{Hoyle05, Rojas05, Patiri06, Douglass19, Habouzit20}.

A number of different types of algorithms have been used to detect voids in the 
cosmic web.  The VoidFinder algorithm \citep{ElAd97, Hoyle02} finds relatively 
empty spheres in the distribution of galaxies and combines them into individual 
voids.  Another common strategy is to link low-density regions together using a 
watershed algorithm.  Several implementations using a Voronoi tessellation to 
approximate local density exist \citep[e.g.][]{Neyrinck08, Sutter15, Nadathur19}.

While the aforementioned algorithms find voids geometrically, voids are expected 
to be dynamically unique structures \citep{Sheth04}.  This dynamical information 
is not typically available in galaxy redshift surveys, but with simulations, it 
is possible to define voids relative to the dynamics of the matter within them.  
Cosmological simulations have been used to predict properties of voids 
\citep{Ricciardelli14, Hamaus14a} and forecast the cosmological constraining 
power of voids \citep{Pisani15}.  One method for identifying dynamical voids in 
simulations is the computation of the number of dimensions along which matter 
has gravitationally collapsed \citep{Falck12}.  As described in 
Section~\ref{sec:origami}, voids are defined as regions undergoing no 
shell-crossing.

In this work, we investigate the ability of geometrical void-finding algorithms 
to detect dynamical void regions.  We apply the ORIGAMI algorithm 
\citep{Falck12} to an $N$-body simulation from the AbacusSummit simulation suite 
\citep{Maksimova21} to determine each dark matter particle's crossing number --- 
the number of dimensions along which it undergoes shell-crossing (the number of 
dimensions of gravitational collapse).  Then, using the crossing numbers to 
classify particles as belonging to voids, walls, filaments, or clusters, we 
study the overlap of these categories with the voids found in the corresponding 
halo catalog identified by two different void-finding algorithms in the Void 
Analysis Software Toolkit \citep[VAST;][]{Douglass22}: VoidFinder and \Vsquared.  
This allows us to quantify the relative accuracy of these algorithms in 
detecting dynamically distinct regions, i.e. regions dominated by low crossing 
numbers, of the large scale structure.

The paper is organized as follows.  In Section~\ref{sec:theory}, we describe the 
theory of void evolution relevant to our study, and in Section~\ref{sec:sim}, we 
describe the AbacusSummit simulation suite and the properties of the simulation 
used in the crossing number analysis.  Sections~\ref{sec:origami} and 
\ref{sec:vast} present the algorithms used to compute crossing numbers of dark 
matter particles and to find voids in the distribution of halos, respectively.  
Results are discussed in Section~\ref{sec:results}, examining the relationships 
between void regions defined by void-finding algorithms and the void particles 
identified by ORIGAMI.  In Section~\ref{sec:poor}, we explore ways to mitigate 
the relatively poor classification of the watershed algorithm, \Vsquared, and in 
Section~\ref{sec:conclusion}, we summarize our results.


\section{Excursion Formalism}\label{sec:theory}

The excursion-set formalism, first proposed by \cite{Press74}, provides an 
analytical model of the gravitational collapse and virialization of dark matter 
halos.  \cite{Bond91} developed the excursion model to describe the halo mass 
function, but the model was not applied to a theoretical description of voids 
until the pioneering work of \cite{Sheth04}.

The excursion-set model of voids describes the evolution of an initial 
underdensity in the matter distribution.  \cite{Sheth04} find that, as matter is 
attracted to the overdense surroundings, the typical shape of underdensities 
becomes more spherical and can be effectively described by a series of spherical 
shells.  More central shells experience a stronger ``repulsion'' from void 
centers, leading to a distinct ``wall'' feature --- a build up of expanding 
matter near the void edge as shells approach and cross each other.  We expect, 
then, to find a high but narrow preference for high dark matter particle 
crossing numbers (the number of dimensions along which they have undergone 
shell-crossing) near void edges, along with the preference for low crossing 
numbers in voids' more central regions.

It is also possible to use the excursion-set formalism to model the theoretical 
abundance of spherical underdensities in the Universe.  While we expect good 
agreement between low crossing numbers and void regions due to the dynamical 
processes involved in void formation, crossing numbers are not useful for 
identifying the distinct spherical underdensities whose abundance is described 
by \cite{Sheth04}.

\section{Simulations}\label{sec:sim}

The AbacusSummit simulation suite \citep{Maksimova21} is a set of over 100 
periodic $N$-body simulations spanning several box sizes up to 
$(7.5\text{ \hGpc})^3$, particle mass resolutions down to 
$\sim3\times10^8 M_\odot$, and implementing several cosmologies.  AbacusSummit 
was produced with the Abacus $N$-body algorithm \citep{Garrison21} at the Oak 
Ridge Leadership Computing Facility's Summit supercomputer and is the largest 
high-accuracy $N$-body data set produced to date.

We use an AbacusSummit Hugebase simulation, which evolves $2304^3$ dark matter 
particles of mass $\sim5\times10^{10}M_\odot$ within a periodic box of side 
length 2~\hGpc.  This simulation uses Planck 2018 flat $\Lambda$CDM cosmology: 
$\Omega_m=0.315$, $h=0.674$, $\sigma_8=0.811$, $n_s=0.965$ \citep{Aghanim20}.  
In addition to full particle timeslices, the AbacusSummit data products include 
halo catalogs produced using the CompaSO halo-finder \citep{Hadzhiyska21}.  We 
run all void-finding algorithms on the corresponding halo catalog of the 
Hugebase simulation, which contains $\sim1.8\times10^7$ halos.

\section{Crossing Numbers Using ORIGAMI}\label{sec:origami}

The gravitational collapse of the matter distribution in these $N$-body 
simulations results in the crossing of dark matter particle positions as their 
dynamics become nonlinear.  The number of dimensions of shell-crossing, and thus 
of nonlinearity, determines the classification of dark matter particles into 
various types of large-scale structures: clusters from three-dimensional 
collapse, filaments from two-dimensional collapse, and walls from 
one-dimensional collapse.  Particles which have undergone no shell-crossing are 
defined as belonging to cosmic voids.  The ORIGAMI algorithm \citep{Falck12} 
computes the number of shell-crossing dimensions, referred to as the crossing 
number (CN), by comparing the final relative positions of pairs of particles 
with their relative positions on the initial grid.

Running the base ORIGAMI algorithm on simulations with more than $512^3$ 
particles requires a prohibitively large amount of memory.  We introduce 
modifications that compute crossing numbers for subvolumes of periodic 
simulations by implementing a buffer zone and subsequently merge the subvolume 
results, allowing ORIGAMI to be run on larger simulations.  Crossing numbers in 
the AbacusSummit simulations were computed using this modified version of 
ORIGAMI\footnote{The modified ORIGAMI is available for download at \url{https://github.com/dveyrat/origami/tree/subdivide}.}.

\section{VAST}\label{sec:vast}

\begin{figure*}
  \centering
  \includegraphics[width=0.485 \textwidth]{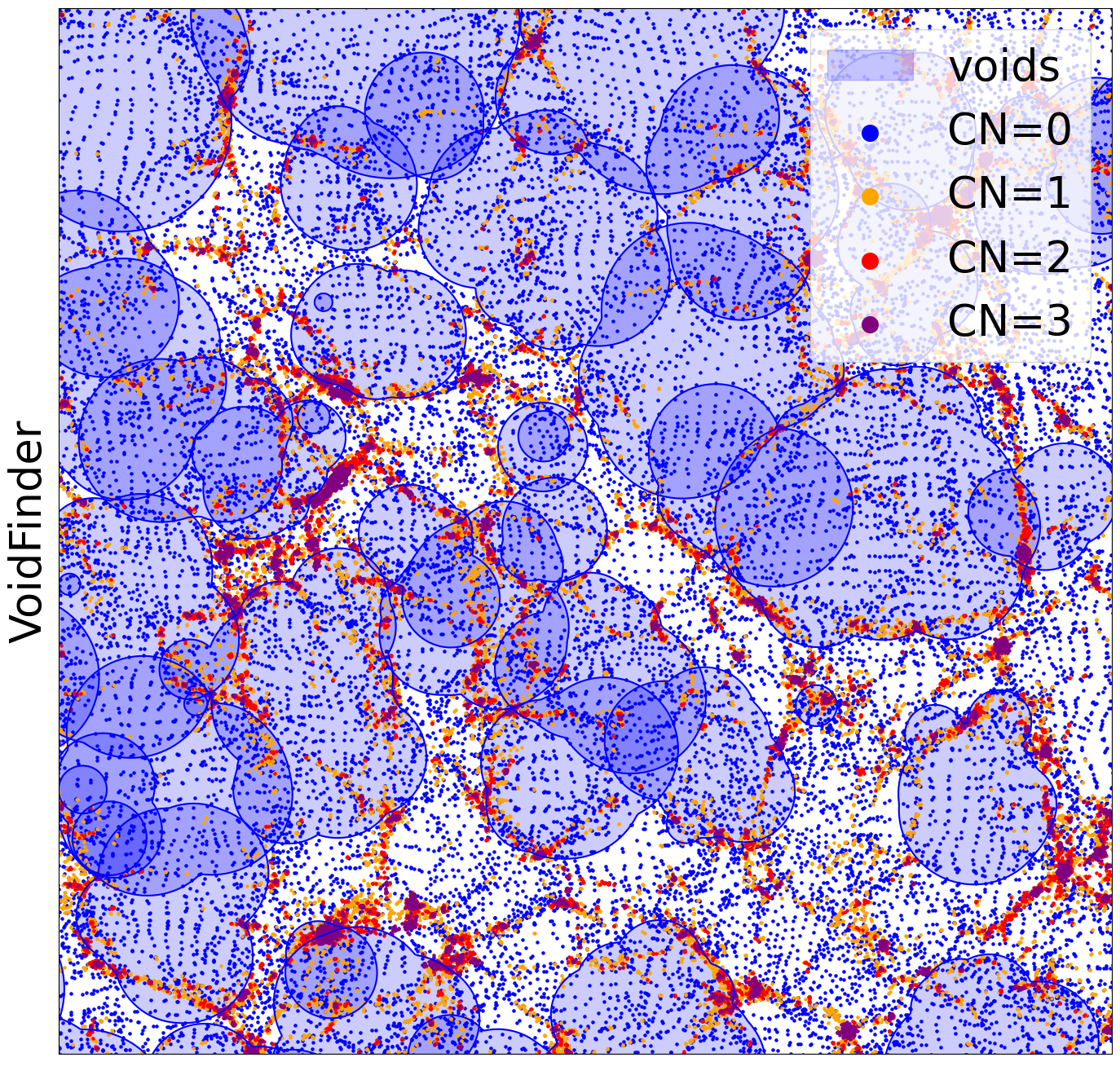}
  \includegraphics[width=0.485 \textwidth]{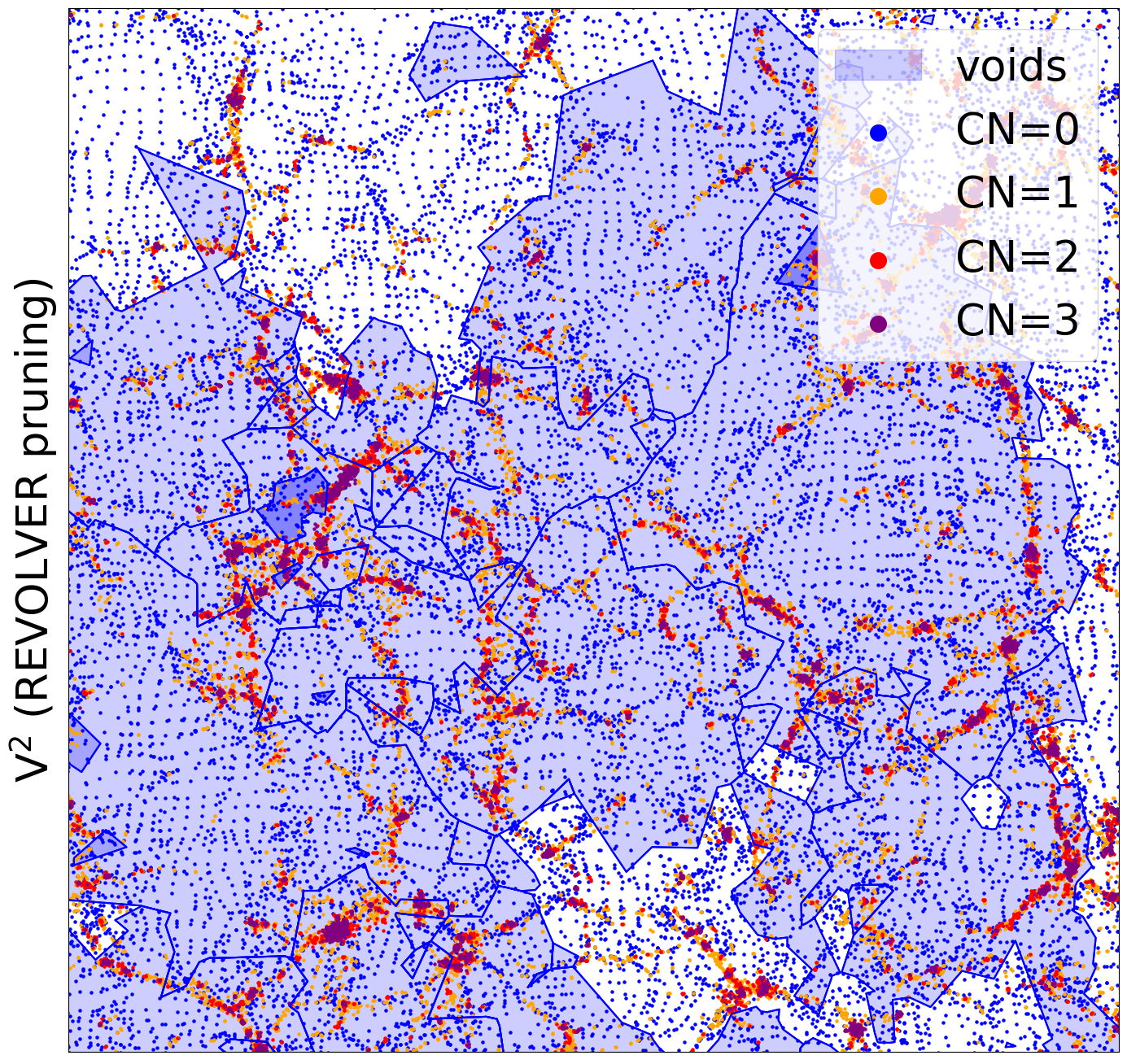} \\
  \includegraphics[width=0.485 \textwidth]{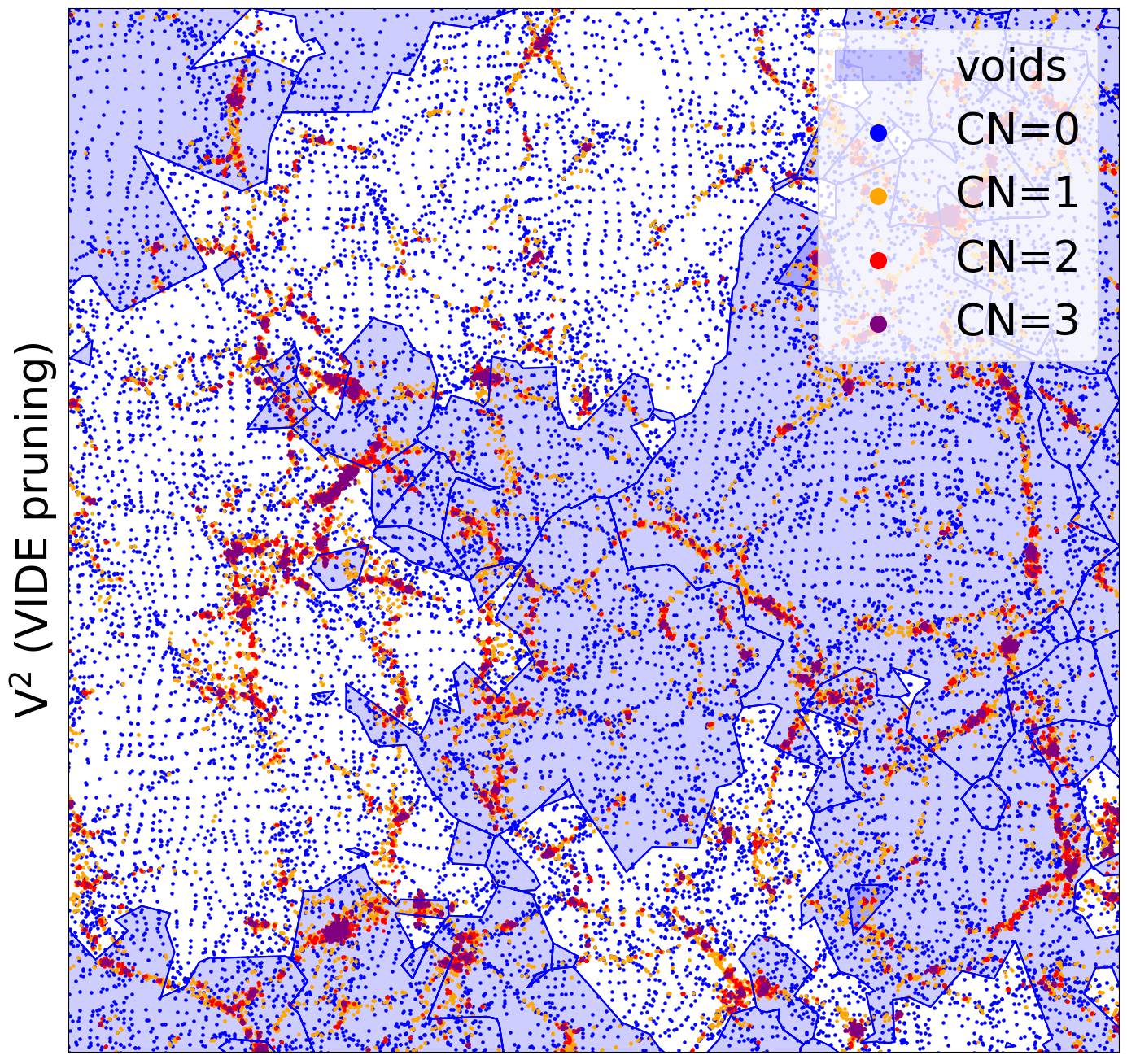}
  \includegraphics[width=0.485 \textwidth]{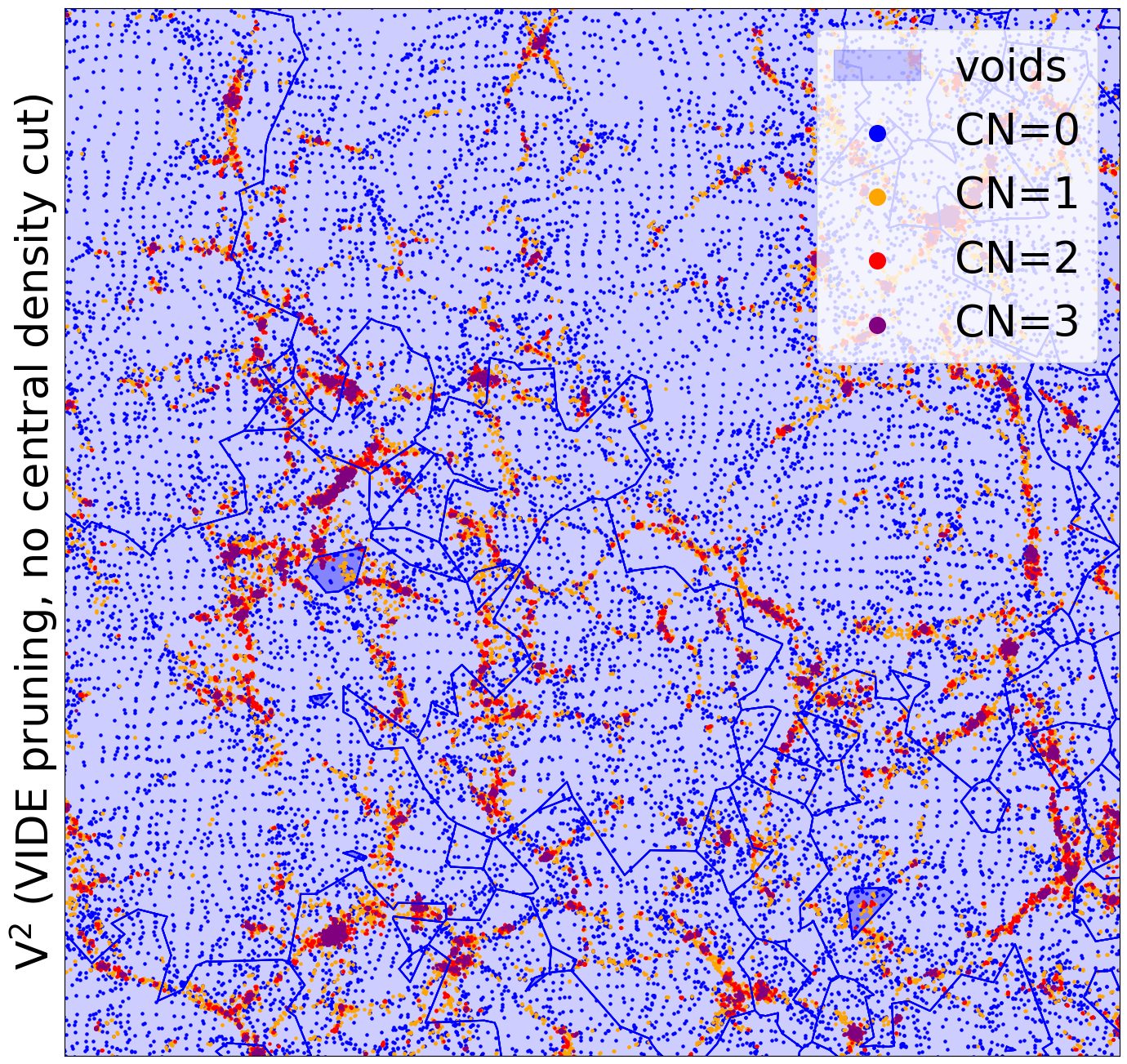}
  \caption{A 100~\hMpc$\times$100~\hMpc subsection of a 10~\hMpc-thick slice of 
  the AbacusSummit Hugebase simulation.  The locations of particles with 
  crossing number 0 are shown in blue, 1 in orange, 2 in red, and 3 in violet.  
  The intersections of voids with the center of the slice are shown in blue for 
  VoidFinder (top left), \Vsquared/REVOLVER (top right), and \Vsquared/VIDE 
  (bottom).}
  \label{fig:slice}
\end{figure*}

We identify voids in the distribution of halos using the Void Analysis Software 
Toolkit\footnote{VAST is available for download at 
\url{https://github.com/desi-ur/vast}.} \citep[VAST;][]{Douglass22}.  VAST 
includes an implementation of the VoidFinder algorithm, which grows and merges 
nearly empty spheres, and \Vsquared, an implementation of the ZOBOV algorithm 
\citep{Neyrinck08}, which uses a watershed method to identify voids.  The voids 
found by each algorithm are shown in Figure~\ref{fig:slice}, overlaying the dark 
matter particles within a 10~\hMpc-thick slice of the AbacusSummit Hugebase 
simulation.

\subsection{VoidFinder}\label{sec:VF}

VoidFinder, originally described by \cite{ElAd97}, begins by removing isolated 
tracers from the catalog, defined as those with a distance to their 
third-nearest neighbor $d_{\rm 3NN} > \overline{d_{\rm 3NN}} + 
1.5\sigma_{d_{\rm 3NN}}$.  The remaining tracers are placed on a 
three-dimensional grid, and a sphere is grown from each empty grid cell until 
its surface is bounded by four tracers.  Next, maximal spheres, defined as 
spheres with radii $r \geq 10$~\hMpc that do not overlap a larger maximal sphere 
by more than 10\% of their volume, are selected from the population of spheres.  
Each maximal sphere is identified as belonging to a unique void, and that void 
is built from the union of all spheres which intersect exactly one maximal 
sphere by at least 50\% of their volume.  See \cite{Hoyle02} for a more detailed 
description of VoidFinder.

\subsection{Voronoi Voids (\Vsquared)}\label{sec:V2}

Voronoi Voids, or \Vsquared, is a Python implementation of the ZOBOV algorithm, 
first described in \cite{Neyrinck08}.  \Vsquared begins by creating a Voronoi 
tesselation of the tracer distribution, whose cell volumes are used as estimates 
of the local density.  A watershed algorithm is used to combine groups of 
adjacent Voronoi cells into ``zones,'' where each cell is put into the same zone 
as its least dense neighboring cell, and any cell which is less dense than any 
of its neighbors is identified as a local density minimum that serves as its 
zone's central cell.

Adjacent zones are linked together by the least-dense pair of adjacent cells 
between them, whose density is referred to as the ``linking density.''  The 
collections of linked zones (including single zones) are identified as voids, 
creating a hierarchy of voids up to one containing the entire catalog.  Several 
methods exist to prune the full hierarchy and extract meaningful voids 
\citep[see][]{Neyrinck08, Sutter15, Nadathur19}.  We use two pruning methods: 
VIDE pruning \citep[\Vsquared/VIDE,][]{Sutter15}, which sets a maximum linking 
density for the zone-linking step, an optional maximum central density for each 
void, and a minimum void radius (we use $10$~\hMpc in this work); and REVOLVER 
pruning \citep[\Vsquared/REVOLVER,][]{Nadathur19}, which labels each zone as a 
unique void.  To our implementation of REVOLVER we add the option to remove any 
voids with effective radii less than the median void effective radius.

There are several methods for computing the centers of ZOBOV voids.  For most of 
the pruning methods, the default void center definition in \Vsquared is the 
barycenter of the tracers within the void, weighted by the volumes of their 
Voronoi cells.  The default for the \Vsquared/REVOLVER pruning algorithm is the 
center of the largest empty sphere which can be inscribed in the void.  In our 
tests of \Vsquared/REVOLVER, we use both of these definitions of void centers.


\section{Results}\label{sec:results}

\begin{deluxetable*}{p{0.3\textwidth}p{0.14\textwidth}p{0.14\textwidth}p{0.14\textwidth}p{0.14\textwidth}}
    \tablewidth{0pt}
    \tablecaption{Crossing number distributions\label{tab:cn0}}
    \startdata
    \tablehead{\colhead{Crossing Number:} & \colhead{0} & \colhead{1} & \colhead{2} & \colhead{3}}
    all DM particles & \hfil 30.8\% & \hfil 19.4\% & \hfil 18.2\% & \hfil 31.6\% \\
    particles in VoidFinder voids & \hfil 46.9\% & \hfil 22.4\% & \hfil 15.6\% & \hfil 15.0\% \\
    particles not in VoidFinder voids & \hfil 16.6\% & \hfil 16.6\% & \hfil 20.5\% & \hfil 46.3\% \\
    particles in \Vsquared/VIDE voids & \hfil 30.6\% & \hfil 19.4\% & \hfil 18.2\% & \hfil 31.8\% \\
    particles not in \Vsquared/VIDE voids & \hfil 31.3\% & \hfil 19.4\% & \hfil 18.1\% & \hfil 31.2\% \\
    particles in \Vsquared/VIDE voids,\linebreak no central density cut & \hfil 30.6\% & \hfil 19.3\% & \hfil 18.2\% & \hfil 31.9\% \\
    particles not in \Vsquared/VIDE voids,\linebreak no central density cut & \hfil 34.4\% & \hfil 19.9\% & \hfil 17.7\% & \hfil 28.1\% \\
    particles in \Vsquared/REVOLVER voids & \hfil 31.3\% & \hfil 19.4\% & \hfil 18.1\% & \hfil 31.2\% \\
    particles not in \Vsquared/REVOLVER voids & \hfil 28.7\% & \hfil 19.1\% & \hfil 18.5\% & \hfil 33.6\% \\
    \enddata
    \tablecomments{Distribution of crossing numbers inside and outside of voids 
    found by VoidFinder and \Vsquared, as well as the total distribution in the 
    AbacusSummit Hugebase simulation.  We report the fraction of particles in a 
    region with a given crossing number.  For example, 46.9\% of all particles 
    in VoidFinder voids have CN~$= 0$.  All uncertainties are below 0.1\%, 
    estimated using several AbacusSummit Hugebase realizations.}
\end{deluxetable*}

\begin{deluxetable}{CCCC}
  \tablewidth{0pt}
  \tablecaption{CN~$=0$ particles not in voids\label{tab:cn00}}
  \tablehead{\colhead{VoidFinder} & \colhead{\Vsquared/VIDE} &  \colhead{\Vsquared/VIDE,} &
  \colhead{\Vsquared/REVOLVER}\\
   & & no central & \\
   & & density cut & }
  \startdata
    28.4\% & 36.4\% & 8.3\% & 14.6\% \\
  \enddata
  \tablecomments{Percentage of particles with CN~$=0$ particles not located in 
  any void found by VoidFinder and \Vsquared.  All uncertainties are below 
  0.1\%, estimated using several AbacusSummit Hugebase realizations.}
\end{deluxetable}

\vspace{-2.5em}

A total of 410,852 voids were found in the Abacus Hugebase simulation by 
VoidFinder, 79,721 voids by \Vsquared/VIDE, 81,680 voids by \Vsquared/VIDE with 
no central density cut\footnote{872 voids with $R_{\rm eff} > 80~\hMpc$ were 
removed from this catalog to prevent the inclusion of large supervoids in the 
void hierarchy containing most of the survey.}, and 43,430 voids by 
\Vsquared/REVOLVER.  The fraction of dark matter particles with each crossing 
number in the AbacusSummit Hugebase simulation, and of those in voids defined by 
each of the void-finding algorithms studied, can be found in 
Table~\ref{tab:cn0}.  Throughout the entire simulation, we find that 31\% of the 
dark matter particles have undergone no shell-crossing, 19\% have undergone 
shell-crossing along one dimension, 18\% along two dimensions, and 32\% along 
three dimensions.  These results are in good agreement with the crossing number 
distributions found by \cite{Falck14} for similar particle resolutions.  Our 
results, using a simulation with an initial grid spacing 
$L/N = (1000$~\hMpc)$/1152$, are similar to those in \cite{Falck14} for initial 
grid spacing $L/N = (100$~\hMpc)$/128$ but with a shift towards more void 
(CN$=0$) particles and fewer cluster (CN$=3$) particles, agreeing with the trend 
for greater initial grid spacing.

\subsection{Crossing Number Distribution within Voids}

\begin{figure*}
  \centering
  \includegraphics[width=0.485\textwidth]{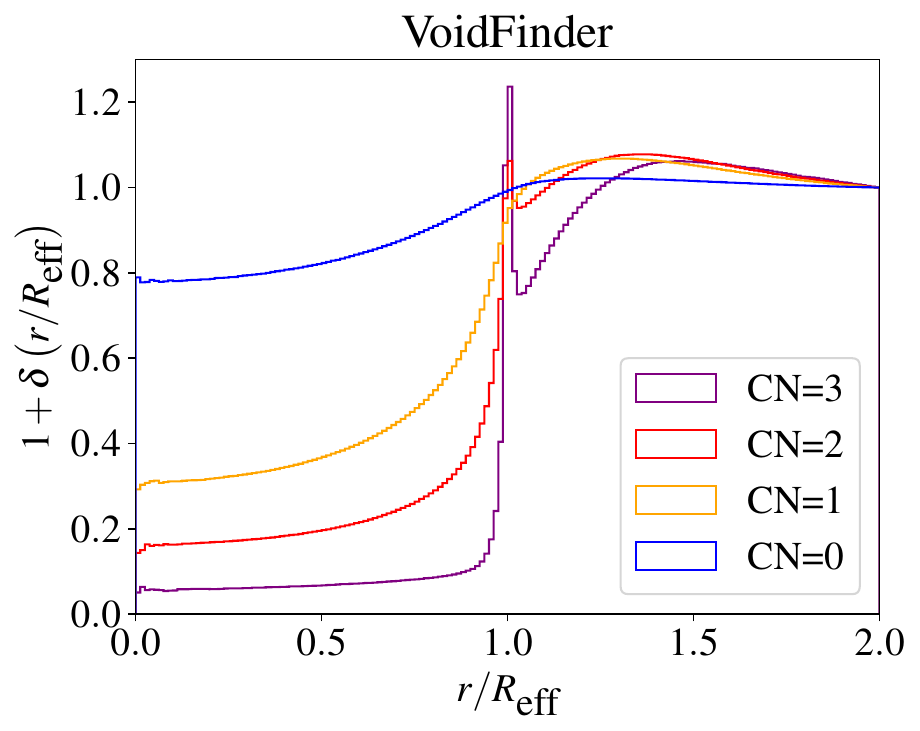} \\
  \includegraphics[width=0.485\textwidth]{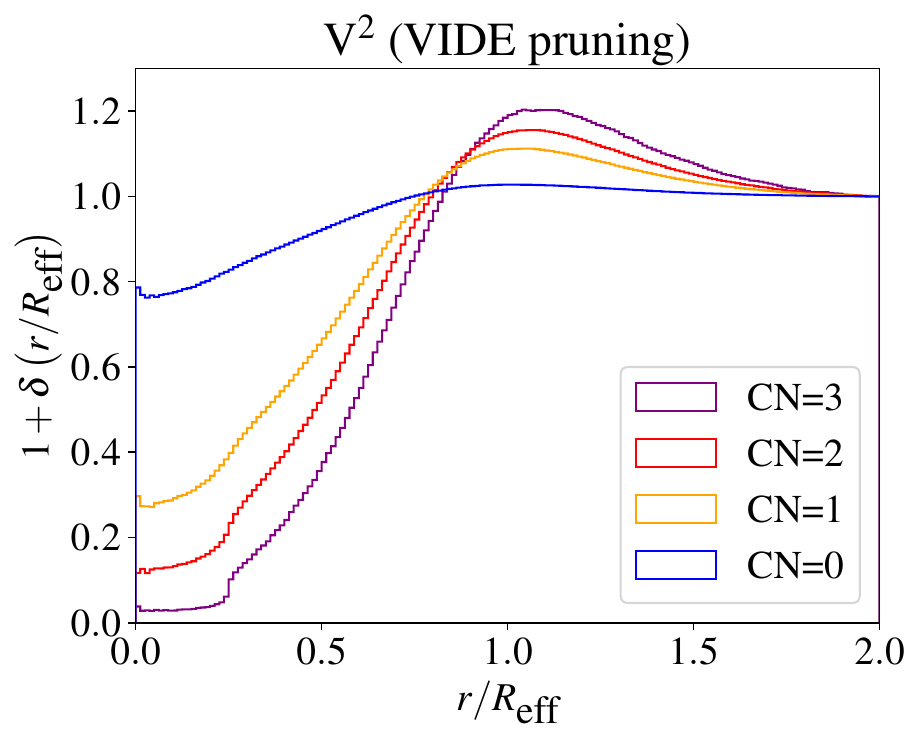}
  \includegraphics[width=0.485\textwidth]{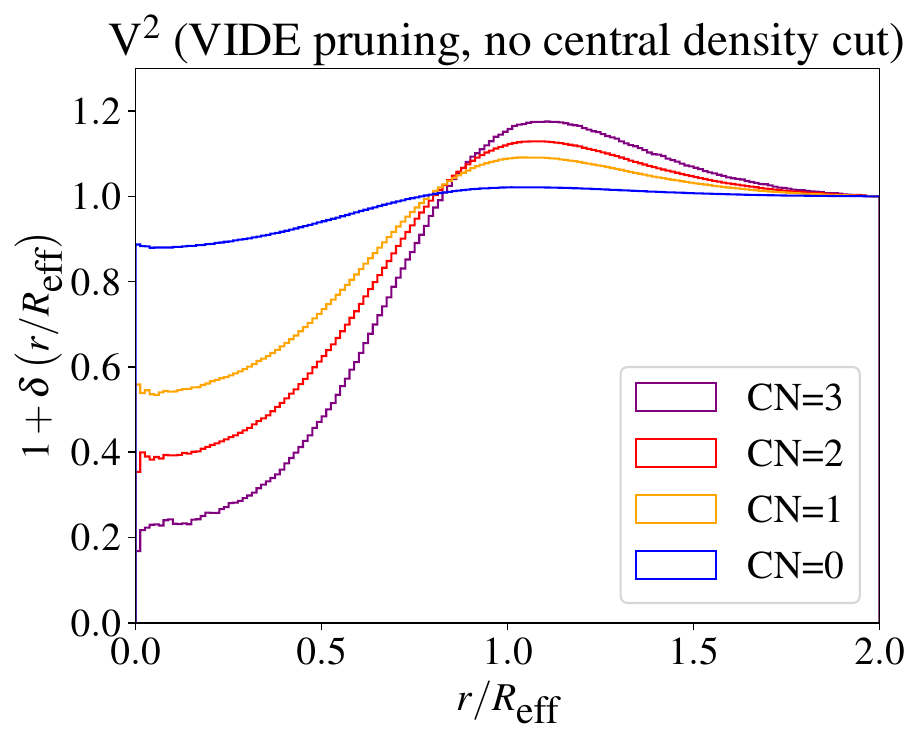} \\
  \includegraphics[width=0.485\textwidth]{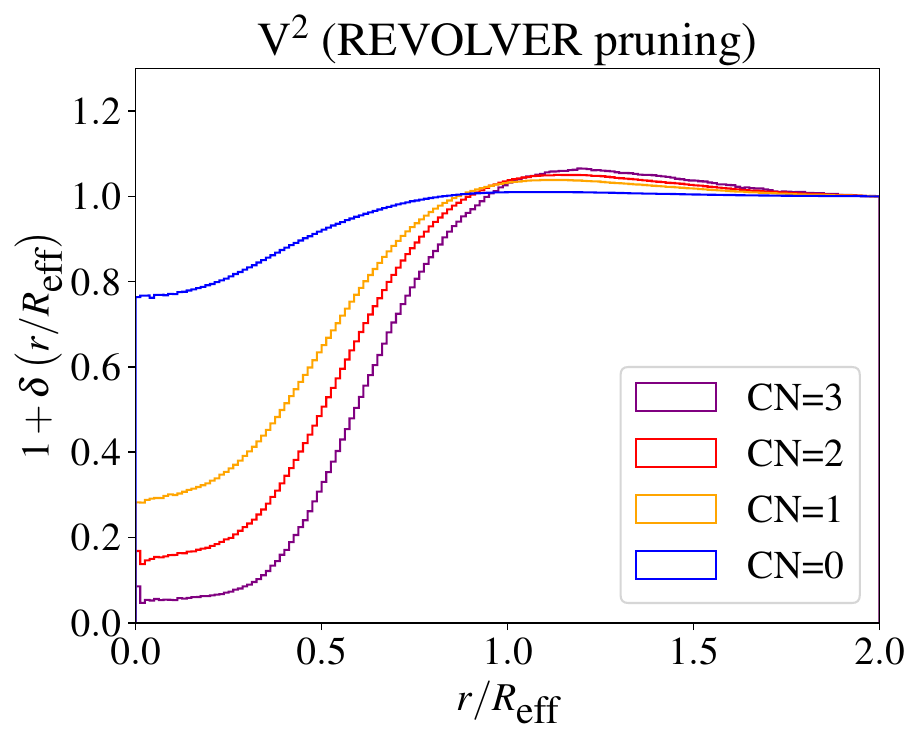}
  \includegraphics[width=0.485\textwidth]{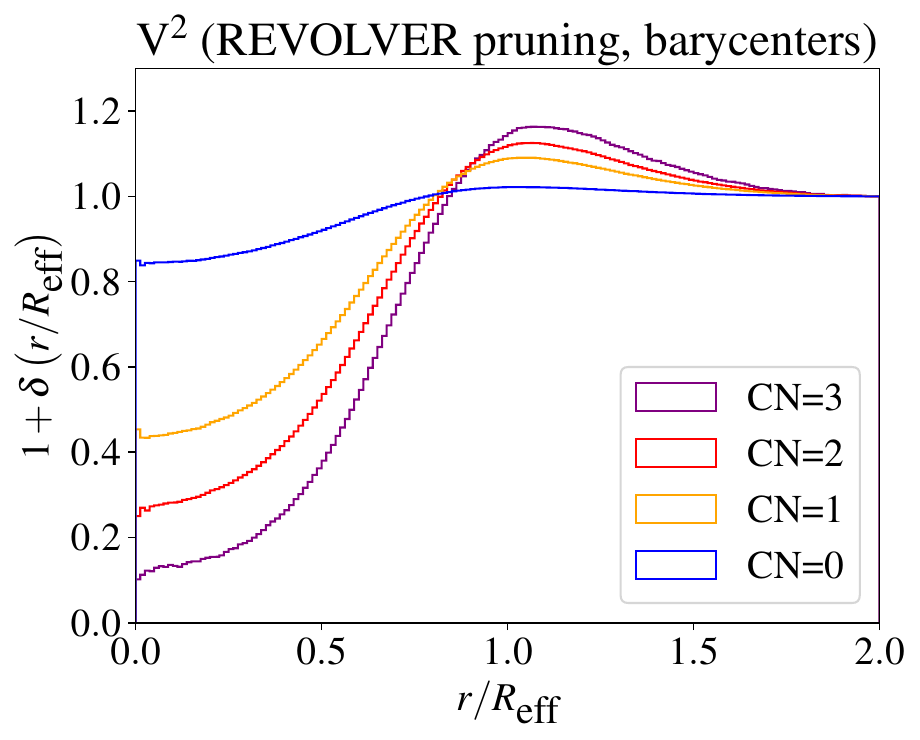}
  \caption{Normalized void density profiles by crossing number for VoidFinder 
  (top), \Vsquared/VIDE (middle), and \Vsquared/REVOLVER (bottom) voids.  All 
  uncertainties are negligibly small, estimated using several AbacusSummit 
  Hugebase realizations.}
  \label{fig:cn2526}
\end{figure*}

\begin{figure*}
  \centering
  \includegraphics[width=0.48\textwidth]{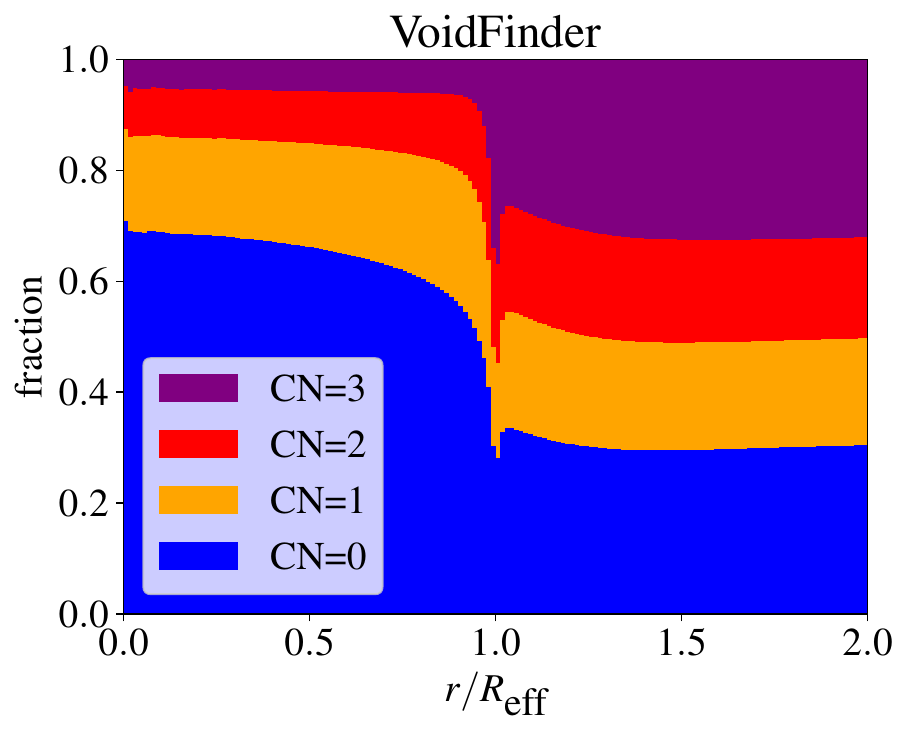} \\
  \includegraphics[width=0.48\textwidth]{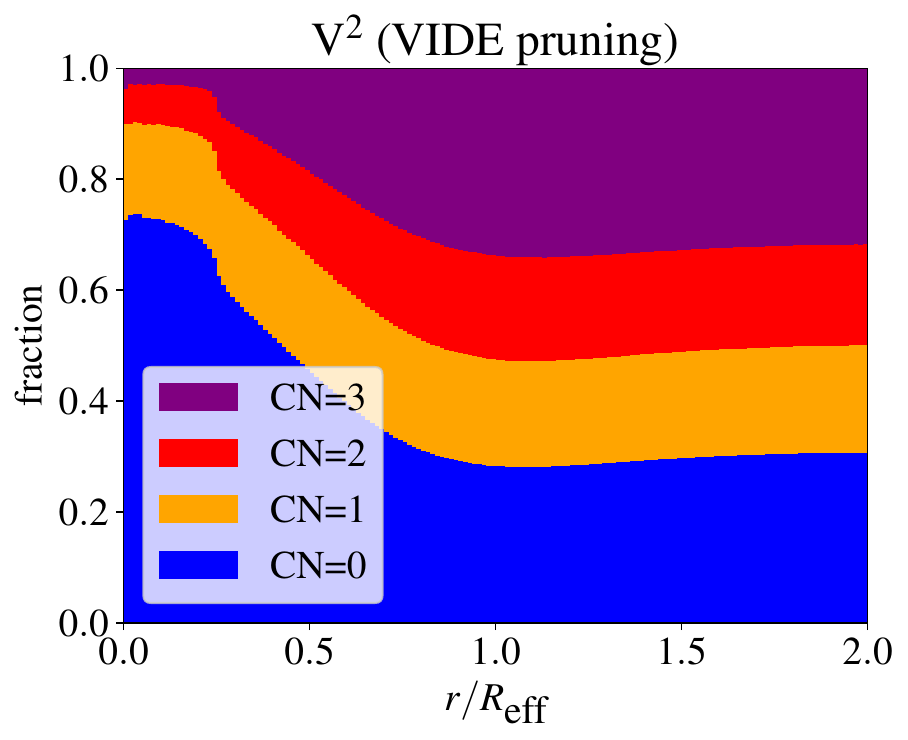}
  \includegraphics[width=0.48\textwidth]{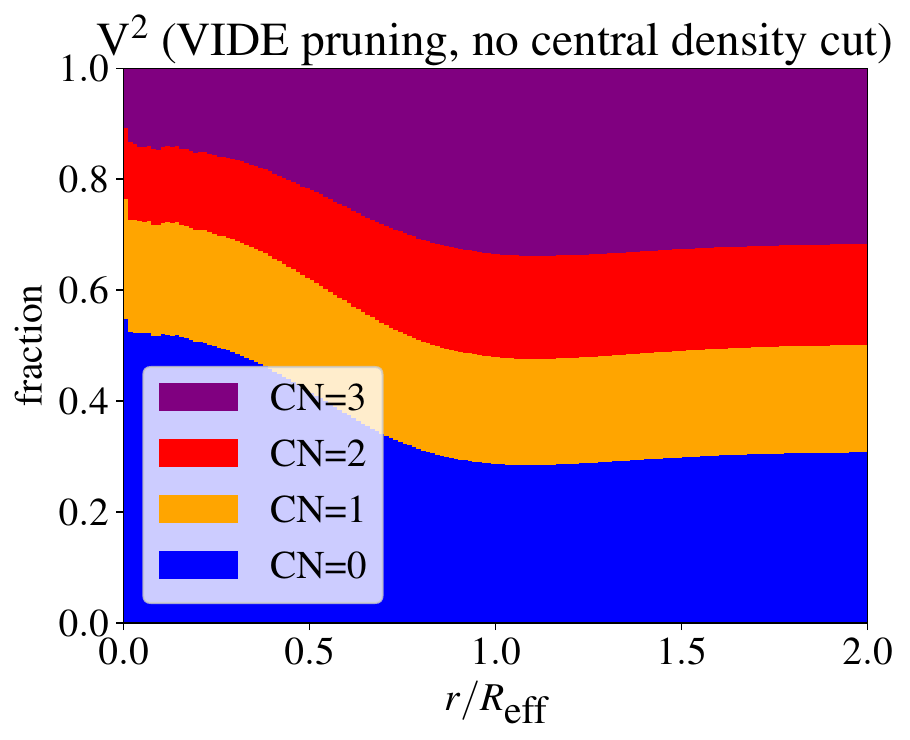} \\
  \includegraphics[width=0.48\textwidth]{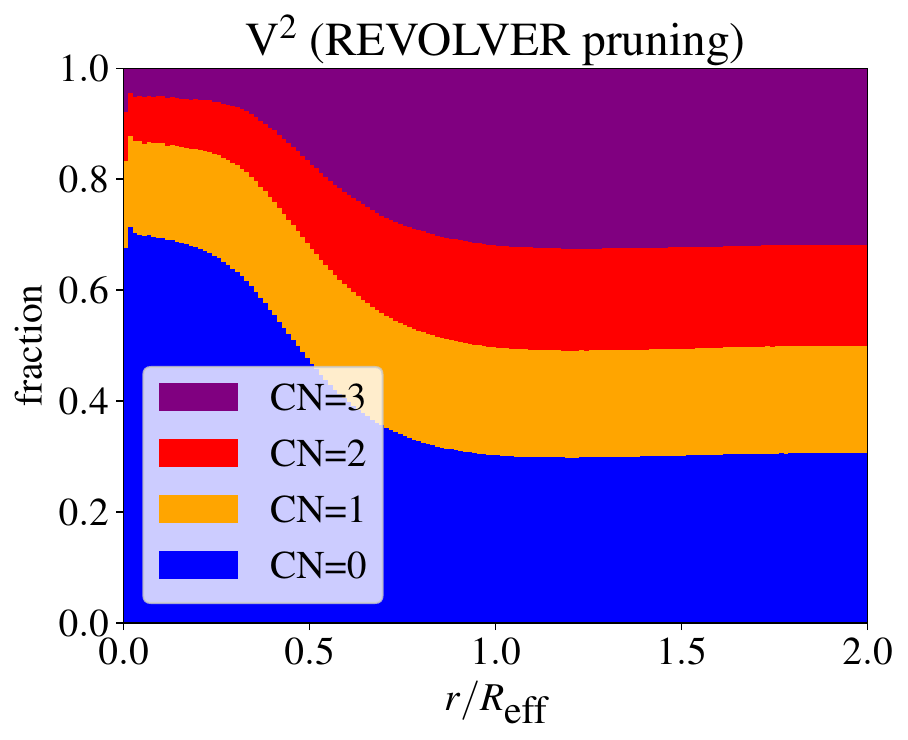}
  \includegraphics[width=0.48\textwidth]{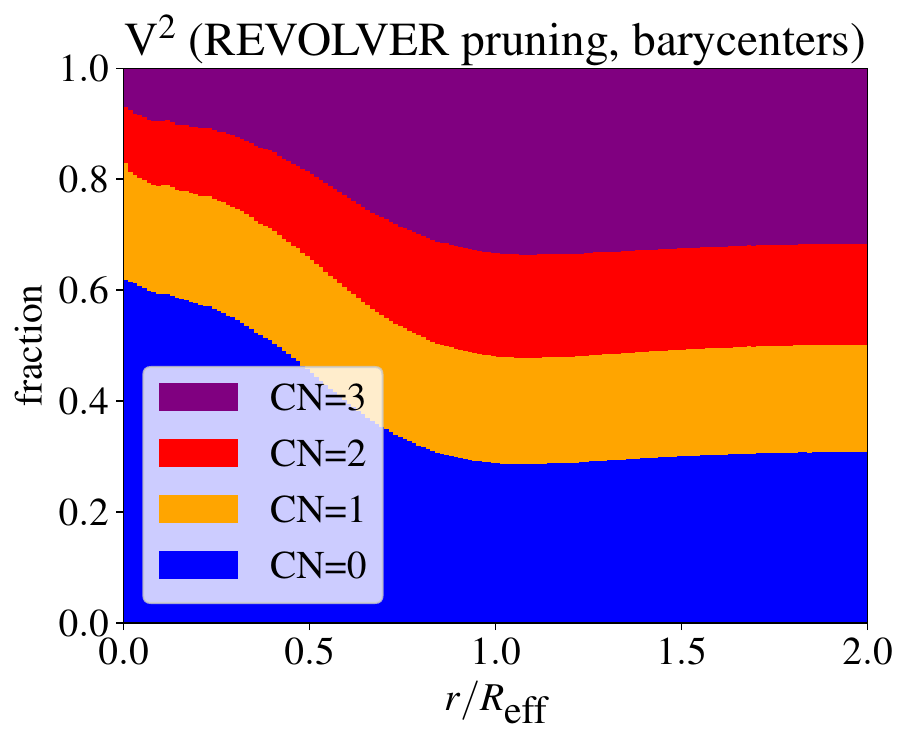}
  \caption{The distributions of crossing numbers by normalized distance from 
  void centers found in real space by VoidFinder (top), \Vsquared/VIDE (middle), 
  and \Vsquared/REVOLVER (bottom).}
  \label{fig:cn2122}
\end{figure*}

We expect the distribution of crossing numbers in void regions to contain an 
excess of CN$=0$ particles because any dark matter particles which are 
identified by ORIGAMI to have undergone shell-crossing are expected to be part 
of a gravitationally bound structure: either wall, filament, or cluster.  To 
determine the effectiveness of VoidFinder and \Vsquared at detecting regions 
with low crossing number, we examine the distribution of crossing numbers inside 
and outside voids.  We observe that the distributions of dark matter particles 
with CN$=0$ (void particles) and CN$=3$ (cluster particles) differ significantly 
when the particles are categorized by whether they are inside or outside a void 
(Table~\ref{tab:cn0}).  The distributions of CN$=1$ and CN$=2$ particles differ 
relatively little between void/non-void regions.  While particles in voids are 
more likely to have CN$=0$ and less likely to have CN$=3$ than the background 
distribution, this preference is significantly stronger in VoidFinder voids, 
where 47\% of particles within the voids have CN$=0$ compared to 31\% in 
\Vsquared/VIDE and \Vsquared/REVOLVER voids, and 31\% overall; only 15\% of 
particles in VoidFinder voids have CN$=3$, compared to 32\% in \Vsquared/VIDE 
voids, 34\% in \Vsquared/VIDE voids with no central density cut, 31\% in 
\Vsquared/REVOLVER voids, and 32\% overall.

Figure~\ref{fig:cn2526} shows the crossing number density profiles of dark 
matter particles around voids, and Figure~\ref{fig:cn2122} shows how the 
distribution of crossing numbers is related to the normalized distance to void 
centers, $r/R_{\rm eff}$, where the effective radius $R_{\rm eff}$ is defined as 
the radius of a sphere with the same volume as the void.  We find that the 
particles closest to void centers are significantly more likely to have crossing 
number 0 than 3 for both VoidFinder and \Vsquared.  However, while this 
preference for low crossing numbers extends to nearly $r = R_{\rm eff}$ for 
VoidFinder voids, the distribution around \Vsquared voids begins to gradually 
shift toward the background crossing number distribution at 
$r \approx 0.25R_{\rm eff}$ for \Vsquared/VIDE voids, and $r \approx 0$ for 
\Vsquared/REVOLVER voids and \Vsquared/VIDE voids with no central density cut.  
At a normalized distance of $r \approx 0.8R_{\rm eff}$, the distribution of 
crossing numbers in \Vsquared voids is nearly identical to the background. 

The expected wall feature of voids is visually evident in both 
Figures~\ref{fig:cn2526} and \ref{fig:cn2122}, but it is more prominent around 
voids found by VoidFinder.  There is a clear, sharp increase in preference for 
higher crossing numbers near $r = R_{\rm eff}$, in good agreement with the 
theory developed in \cite{Sheth04}, which predicts a build-up and crossing of 
mass shells as they expand radially outward from void centers (see 
Section~\ref{sec:theory} for details).  The wall feature around \Vsquared voids 
is both less pronounced and less localized around $r = R_{\rm eff}$.

We also observe a sharp feature at $r = 0.25R_{\rm eff}$ in the distributions of 
crossing numbers in \Vsquared/VIDE voids when the central density cut is 
applied.  This is an artifact of the central density cut, which removes all 
voids with densities above some threshold within $r < 0.25R_{\rm eff}$.  As seen 
in the middle right panel of Figures~\ref{fig:cn2526} and \ref{fig:cn2122}, the 
distribution of crossing numbers around voids found by \Vsquared/VIDE with no 
central density cut lacks this feature.

\subsection{CN=0 Particles not in Voids}

While the majority of particles with crossing number 0 are located within 
VoidFinder voids, 28\% do not fall within a void.  And, while particles with 
lower crossing numbers are more likely to be found within \Vsquared voids than 
those with higher crossing numbers, 36\% of CN$=0$ particles do not fall within 
a \Vsquared/VIDE void, 8\% do not fall within a \Vsquared/VIDE void with no 
central density cut, and 15\% do not fall within a \Vsquared/REVOLVER void. 

To determine whether these particles belong to regions we expect to be 
classified as void, we investigate their local environments.  For each dark 
matter particle, we identify all particles within 2~\hMpc and count the 
particles with a given crossing number.  In a void-like environment, we expect 
fewer neighboring particles with higher crossing numbers, as well as a lower 
density of particles overall, leading to a lower particle count for CN$>0$ 
particles relative to CN$=0$ and a lower total particle count.

\begin{deluxetable}{lccc}
    \tablewidth{0pt}
    \tablecaption{Local environment of CN=0 particles\label{tab:cn1}}
    \startdata
    \tablehead{\colhead{Crossing Number:} & \colhead{0} & \colhead{1--3} & \colhead{total}}
    all & 19 (11, 28) & 11 (0, 61) & 32 (13, 86) \\
    VoidFinder: & \phn & \phn & \phn \\
    -- inside voids & 18 (11, 27) & 7 (0, 41) & 28 (12, 67) \\
    -- outside voids & 21 (13, 30) & 27 (3, 121) & 51 (20, 146) \\
    \Vsquared/VIDE: & \phn & \phn & \phn \\
    -- inside voids & 19 (11, 28) & 11 (0, 61) & 33 (14, 87) \\
    -- outside voids & 19 (11, 28) & 10 (0, 59) & 32 (13, 85) \\
    \multicolumn{3}{l}{\Vsquared/VIDE, no central density cut:} & \phn \\
    -- inside voids & 19 (11, 28) & 11 (0, 62) & 33 (14, 87) \\
    -- outside voids & 18 (10, 27) & 9 (0, 52) & 29 (12, 77) \\
    \Vsquared/REVOLVER: & \phn & \phn & \phn \\
    -- inside voids &  19 (11, 28) &  10 (0, 60) &  32 (13, 86) \\
    -- outside voids &  19 (11, 28) &  12 (0, 65) &  34 (14, 90) \\
    \enddata
    \tablecomments{Median number of particles $<2$~\hMpc from particles with 
    CN$=0$ that fall inside or outside a void.  Values in parentheses indicate 
    the boundaries of the central 68\% of the distribution.}
\end{deluxetable}

\vspace{-1em}

The crossing numbers within 2~\hMpc of CN$=0$ particles inside and outside of 
voids are shown in Table~\ref{tab:cn1} for both VoidFinder and \Vsquared.  
Overall, CN$=0$ particles are found to be located in void-like environments with 
little clustering (CN$>0$ particles).  CN$=0$ particles outside of VoidFinder 
voids, however, tend to be located in environments with more neighboring 
particles overall and several times more CN$>0$ particles than in the local 
environments of CN$=0$ particles in VoidFinder voids.  This suggests that, while 
dynamical information alone classifies all CN$=0$ particles as belonging to void 
regions, many are located in significantly denser environments and subsequently 
do not belong to cosmic voids.  These are therefore expected to not be located 
in VoidFinder voids.

The local environments of CN$=0$ particles in \Vsquared voids, however, are much 
more similar to those of their undetected counterparts and of CN$=0$ particles 
as a whole.  Like the overall crossing number distributions in 
Table~\ref{tab:cn0}, this indicates worse classification of void regions by 
\Vsquared, with no evident environmental distinction between the CN$=0$ 
particles inside voids and those outside.

\subsection{Void Finding in Redshift Space}

We also run both void-finding algorithms on a version of the halo catalog 
modified using peculiar velocity information to simulate the effect of peculiar 
velocities on apparent positions, an effect known as redshift-space distortions 
(RSDs).  Because peculiar velocities within voids tend to be directed outwards, 
voids appear elongated along the line-of-sight in redshift space, and we expect 
a smoothing of features in the radial distribution of crossing numbers around 
voids.

\begin{figure*}
  \centering
  \includegraphics[width=0.48\textwidth]{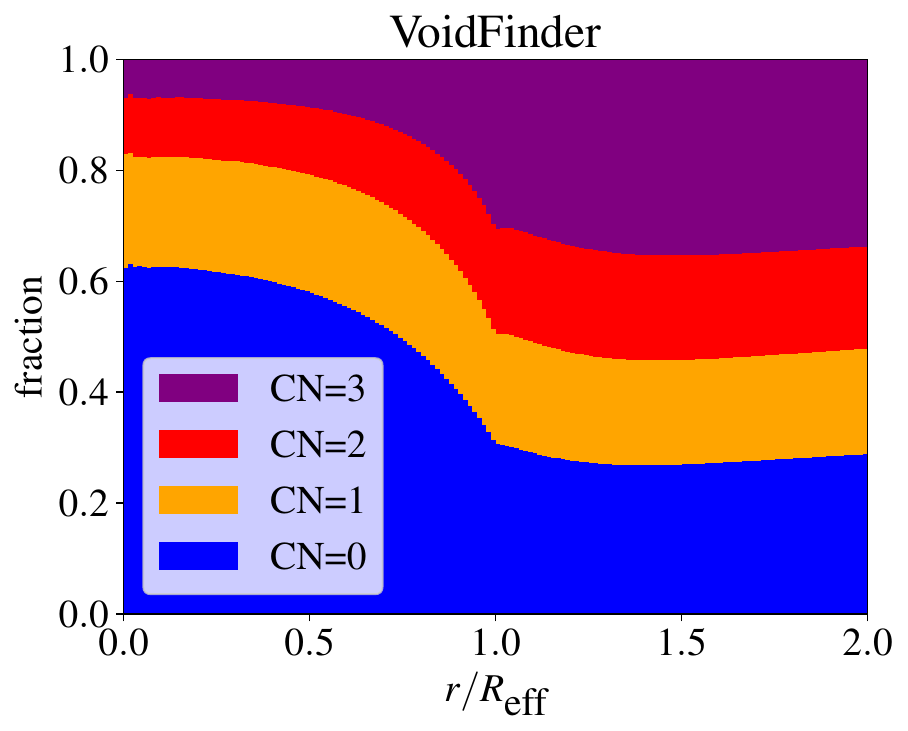} \\
  \includegraphics[width=0.48\textwidth]{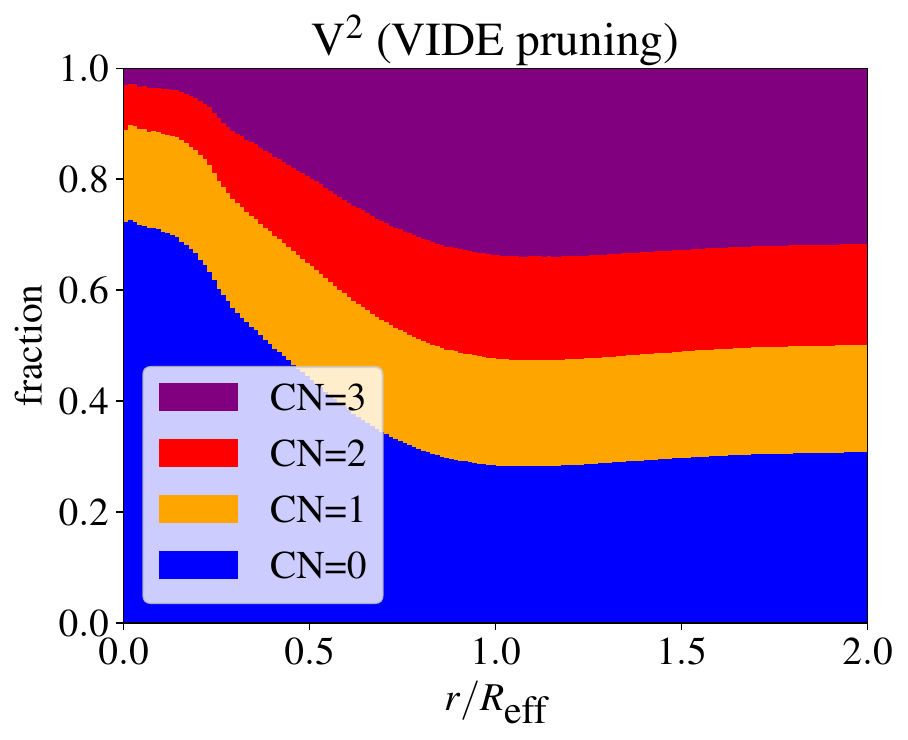}
  \includegraphics[width=0.48\textwidth]{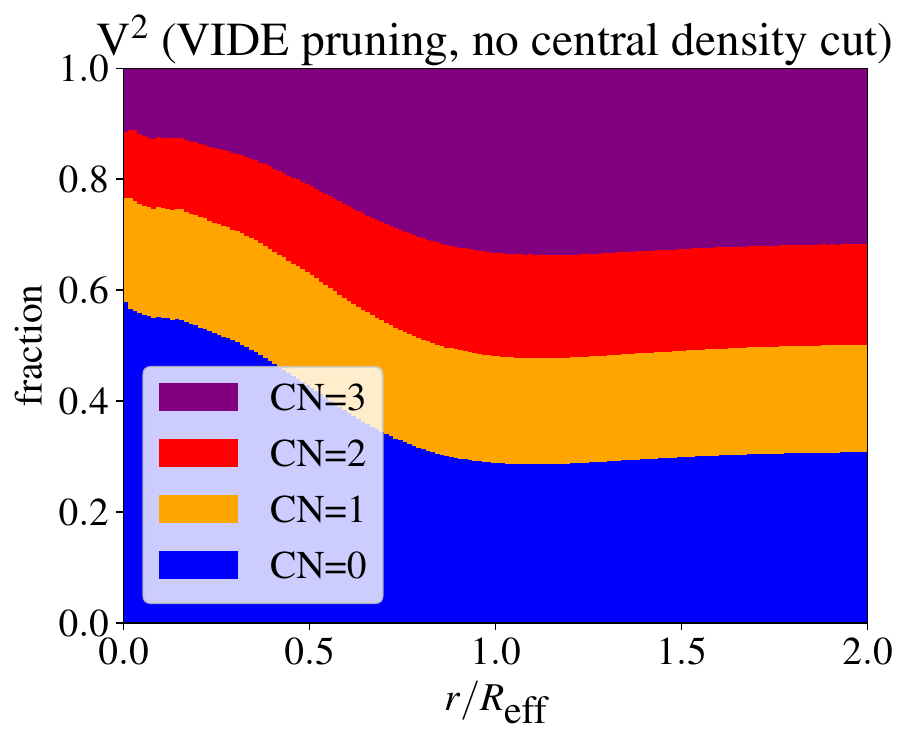} \\
  \includegraphics[width=0.48\textwidth]{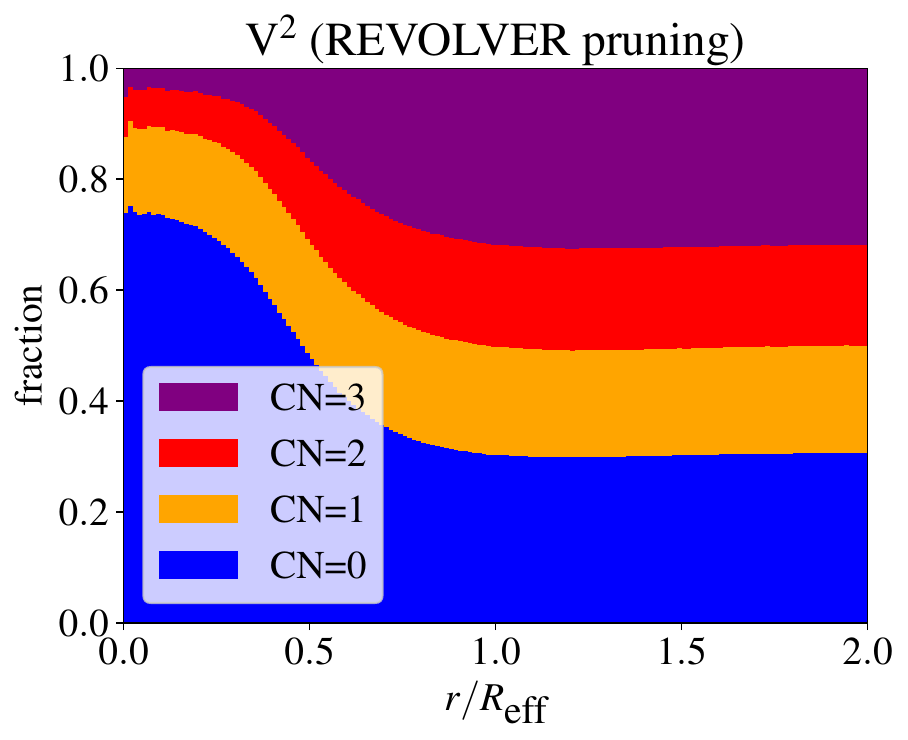}
  \includegraphics[width=0.48\textwidth]{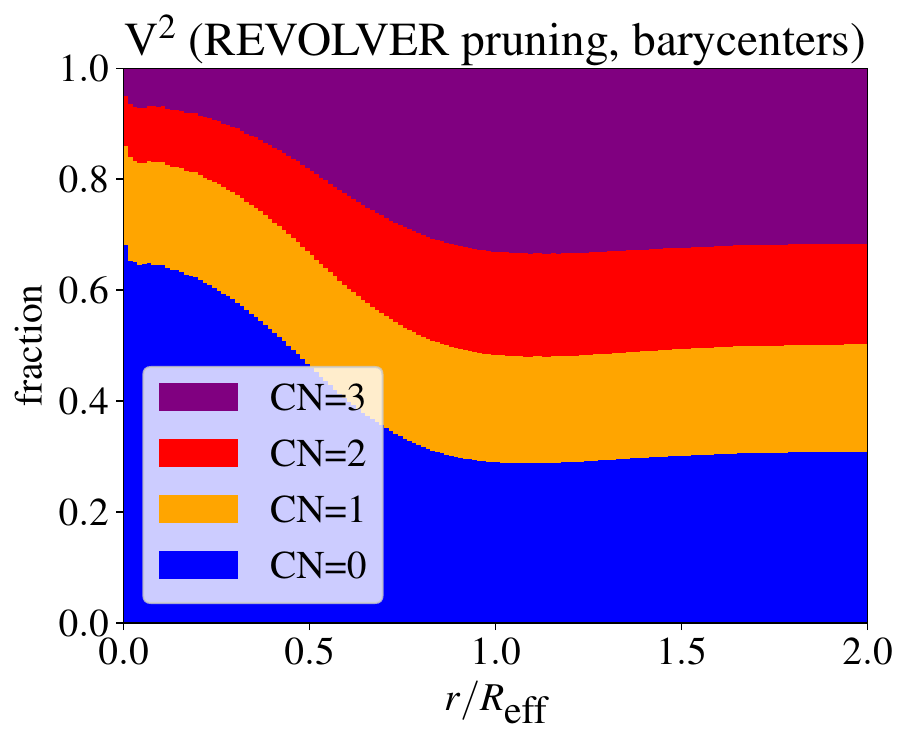}
  \caption{The distributions of crossing numbers by normalized distance from 
  void centers found in redshift space by VoidFinder (top), \Vsquared/VIDE 
  (middle), and \Vsquared/REVOLVER (bottom).}
  \label{fig:cn3940}
\end{figure*}

Figure~\ref{fig:cn3940} shows how the distribution of crossing numbers is 
related to normalized distance to void centers found in redshift space.  While 
the distributions are similar to those shown in Figure~\ref{fig:cn2122} at low 
and high $r/R_{\rm eff}$, the distinct features at $r/R_{\rm eff} = 1$ for 
VoidFinder and $r/R_{\rm eff} = 0.25$ for \Vsquared/VIDE appear to be smoothed 
out, as expected due to the distortion of the voids along the line of sight.  
This suggests that performing reconstruction of real-space positions in a 
redshift catalog before running a void-finding algorithm may improve the voids 
found.

\subsection{Summary of Results}

We expect void regions to contain an excess of particles with CN$=0$ relative to 
the background distribution, and voids found by both VoidFinder and \Vsquared 
contain such an excess.  In \Vsquared voids, however, there is a strong excess 
only in the most central ($r < 0.25R_{\rm eff}$) regions, while in VoidFinder 
voids it extends nearly to $R_{\rm eff}$.  Further, there is an excess of CN$=3$ 
(cluster) particles near the edge of VoidFinder voids, in agreement with 
dynamical theories of void evolution.  We conclude that VoidFinder identifies 
void regions more accurately than \Vsquared.  While this is particularly 
relevant for studies of astrophysics in the void environment, it is less 
indicative of performance in cosmological void analyses, which depend more on 
the identification of void centers than void regions.

\section{Mitigating Poor Classification of \Vsquared voids}\label{sec:poor}

\subsection{Linking Density}

While the \Vsquared algorithm is mostly parameter-free, there is an input 
parameter in \Vsquared/VIDE: the maximum zone-linking density, which limits 
the density of Voronoi cells which can link adjacent zones into voids.  The 
maximum linking density (see Section~\ref{sec:V2} for details) is defined 
relative to the mean number density of the catalog, $\bar{n}$, and is equal to 
$0.2\bar{n}$ by default; it can be as low as 0 (allowing no linking of zones, 
similar to \Vsquared/REVOLVER) or it can be left undefined, leading to all zones 
being linked together into a single void.

While a lower maximum linking density is expected to limit the merging of voids 
across denser regions, the growth of \Vsquared voids up to density maxima occurs 
during the zone creation step, rather than in the subsequent zone-merging step.  
Consequently, varying the amount of zone merging that occurs is not expected to 
mitigate the inclusion of the dense shell-crossing region in the void volume of 
\Vsquared/VIDE voids.

We examine the effect of the maximum linking density on the distribution of 
crossing numbers within \Vsquared/VIDE voids.  The central density cut of the 
\Vsquared/VIDE algorithm is varied to match the maximum linking density.  These 
results are shown in Table~\ref{tab:cn2}.  The percent of particles in voids 
with a given crossing number does not vary by more than $\sim$1\% across 
different linking densities.  This is a minor shift compared to the difference 
between VoidFinder voids and \Vsquared/VIDE voids ($\sim$12\% for both CN$=0$ 
and CN$=3$; see Table~\ref{tab:cn0}), and still leaves the \Vsquared/VIDE 
distribution relatively similar to the distribution of all crossing numbers.

\begin{deluxetable}{C|CCCC}
  \tablewidth{0.49\textwidth}
  \tablecaption{Crossing number distributions\label{tab:cn2}}
  \tablehead{\colhead{crossing} &\multicolumn{4}{c}{maximum linking or central density} \\ \colhead{number} & \colhead{$0.1\bar{n}$} & \colhead{$0.2\bar{n}$} & \colhead{$0.5\bar{n}$} & \colhead{$\bar{n}$}}
  \startdata
    0 & 32.2\% & 33.0\% & 33.2\% & 33.2\% \\
    1 & 20.9\% & 20.9\% & 20.8\% & 20.8\% \\
    2 & 18.5\% & 18.4\% & 18.2\% & 18.2\% \\
    3 & 28.4\% & 27.7\% & 27.8\% & 27.8\% \\
  \enddata
  \tablecomments{Effect of maximum linking or central density on the 
  distribution of crossing numbers within \Vsquared/VIDE voids.  Void-finding 
  was done using a (111~\hMpc)$^3$ subvolume of the AbacusSummit Hugebase 
  simulation.}
\end{deluxetable}

\vspace{-2em}

\subsection{Depth-in-Void}

\begin{figure*}
\centering
  \includegraphics[width=0.475\textwidth]{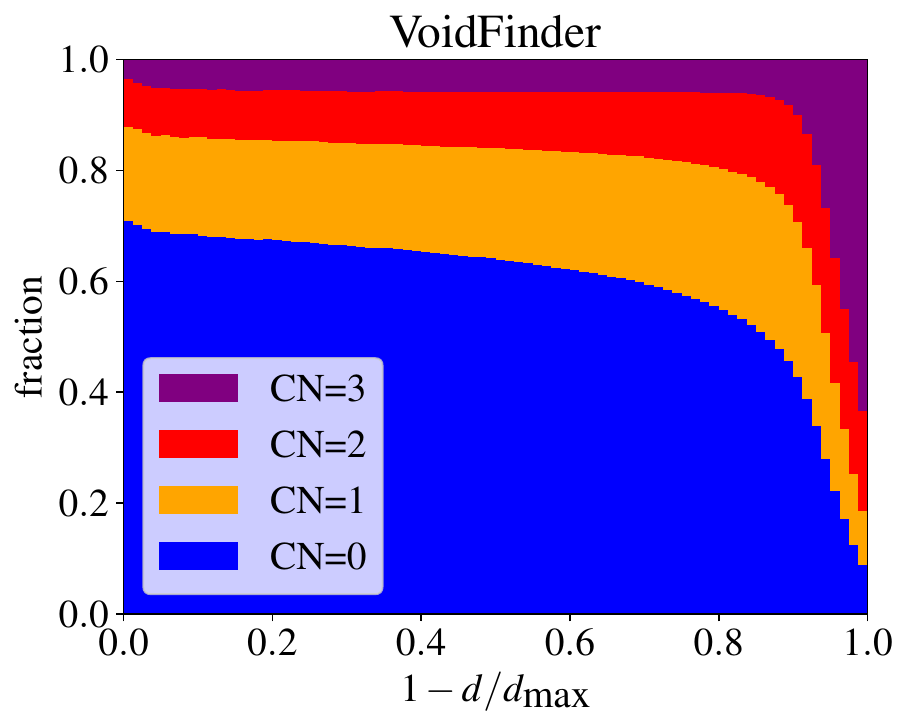}
  \includegraphics[width=0.475\textwidth]{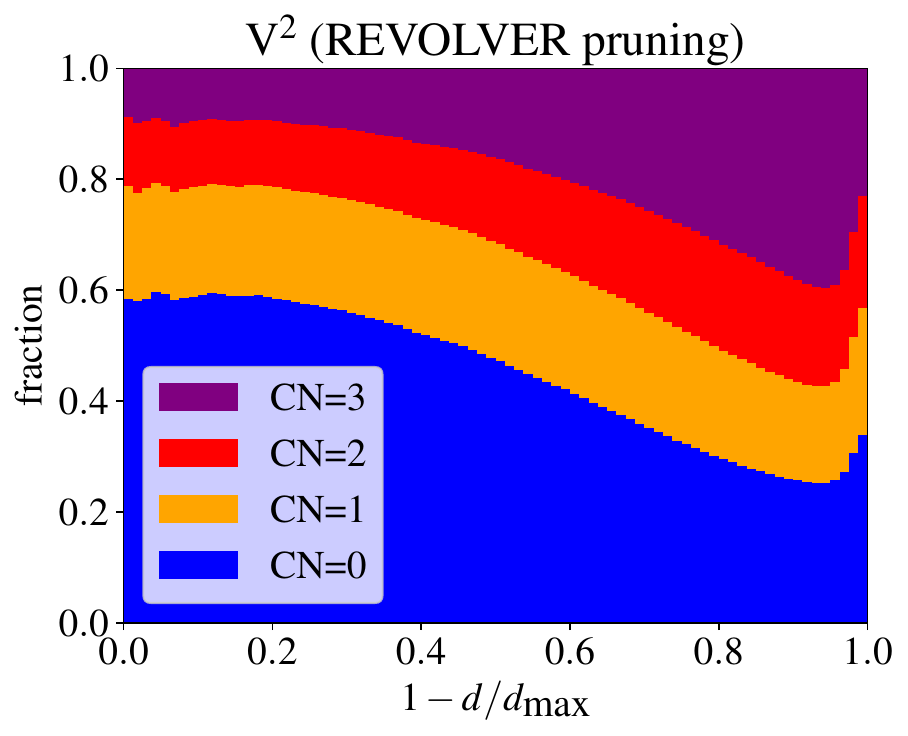} \\
  \includegraphics[width=0.475\textwidth]{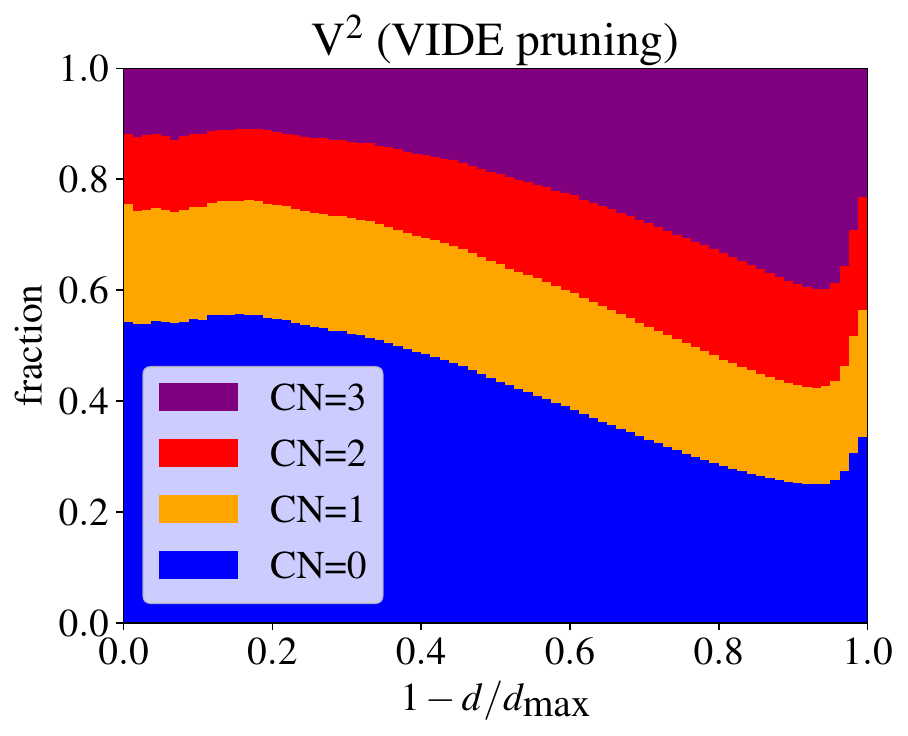}
  \includegraphics[width=0.475\textwidth]{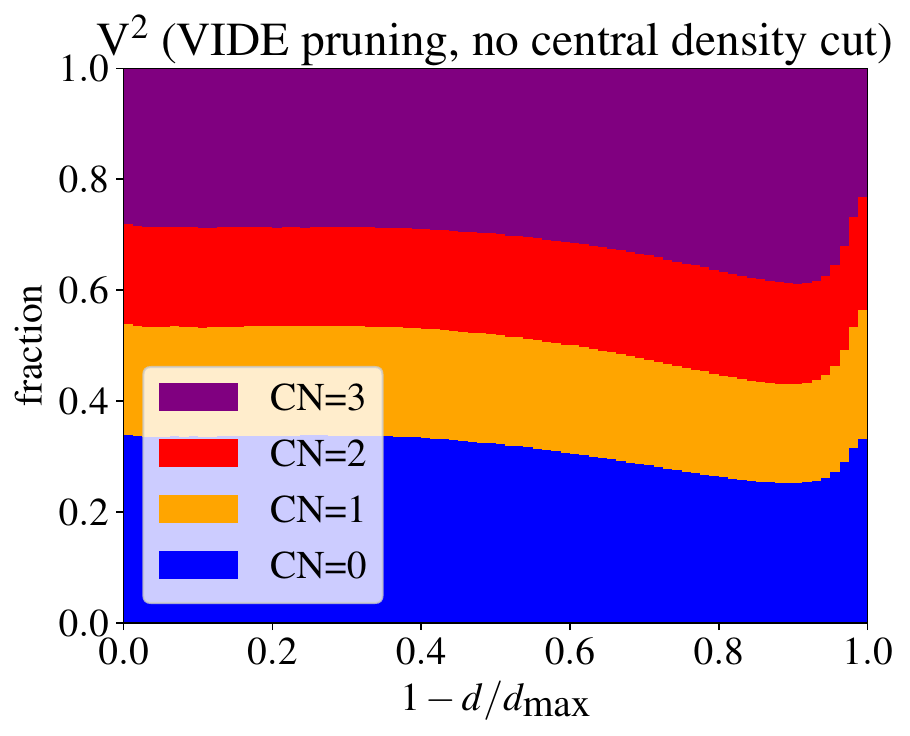}
  \caption{Distribution of crossing numbers by normalized depth from void edges 
  within VoidFinder (top left), \Vsquared/REVOLVER (top right), and 
  \Vsquared/VIDE (bottom) voids.  Depths were computed only for a (1~\hGpc)$^3$ 
  subvolume of the AbacusSummit Hugebase simulation.}
  \label{fig:cn29}
\end{figure*}

While the centers of \Vsquared/VIDE voids have similar crossing number 
distributions to those of VoidFinder voids, this is only out to roughly 
$0.25R_{\rm eff}$ (see the middle panel of Figure~\ref{fig:cn2122}).  
Restricting the analysis to central regions using distance from the void center 
is therefore not an effective modification of \Vsquared, as a significant amount 
of the void volume is omitted, restricting analysis to a small subset of void 
regions. 

Because \Vsquared voids generally have irregular shapes, distance from the edge 
(depth-in-void) may serve as a better measure of centrality than distance from 
the center, which is more effective for the more spherical voids found by 
VoidFinder.  \cite{Zaidouni24} show that void galaxies with a normalized 
distance from the \Vsquared/VIDE void edge of at least $0.4d_{\rm max}$ have a 
distribution of astrophysical properties similar to those of galaxies within 
VoidFinder voids.

We examine the distribution of crossing numbers at different depths within 
\Vsquared voids to determine whether depth is a better measure of centrality 
than distance from the center.  The distance from the void edge is defined as 
the distance to the nearest polygon making up the surface of the void.  The 
results are shown in Figure~\ref{fig:cn29} for both \Vsquared and VoidFinder.  
The previously observed trend of lower crossing numbers for more centrally 
located void particles is apparent, but just as in the distribution based on 
distance from the void center (Figure~\ref{fig:cn2122}), the \Vsquared voids do 
not exhibit the sharp wall feature observed in VoidFinder voids.  Further, the 
central regions of voids defined using depths ($1 - d/d_{\rm max} < 0.25$) do 
not favor low crossing numbers as strongly as central regions defined using 
distance from void centers ($r/R_{\rm eff} < 0.25$).  The preference for low 
crossing numbers is present but particularly weak in the case of \Vsquared/VIDE 
with no central density cut.  These results indicate that there is no clear 
method for extracting the dynamically-distinct parts of the void regions from 
\Vsquared voids.  The shift back to lower crossing numbers at the edge of 
\Vsquared voids can be attributed to the fact that these void edges are made up 
of boundaries between halos' Voronoi cells, an inherently low-density region.

\section{Conclusions}\label{sec:conclusion}

We study the relative accuracy of two different void-finding algorithms in 
detecting dynamical void regions by examining the positions of their voids 
relative to non-clustering regions in the dark matter distribution.  Using a 
(2~\hGpc)$^3$ $N$-body simulation from the AbacusSummit simulation suite 
\citep{Maksimova21}, the evolutionary history of each dark matter particle is 
identified using a modified version of the ORIGAMI algorithm \citep{Falck12}, 
which counts the number of dimensions along which dark matter particles have 
undergone shell-crossing.  Identifying particles with crossing number 0 as 
belonging to voids, we compare their positions with voids found in the 
corresponding halo distribution by VoidFinder and \Vsquared, two void-finding 
algorithms implemented in VAST \citep{Douglass22}.

We find that while both void-finding algorithms produce voids with a central 
bias towards lower crossing numbers, this bias gradually diminishes at greater 
distances from \Vsquared void centers, while the preference remains strong up 
to the edge of VoidFinder voids.  This suggests that \Vsquared includes the 
shell-crossing region surrounding voids as part of the void volume.  Further, at 
the void edge defined as $r/R_{\rm eff} \approx 1$, voids found by VoidFinder 
have a preference for higher crossing numbers that is even stronger than the 
background distribution at larger radii, in good agreement with the 
shell-crossing predictions of the excursion-set formalism by \cite{Sheth04}.  
This feature is absent in the distribution of crossing numbers around voids 
found by \Vsquared.

Given the relative inability of \Vsquared to identify dynamically-distinct void 
regions, we attempt several methods to improve the classification of void 
regions by \Vsquared.  The \Vsquared/VIDE algorithm has one user-set parameter, 
the maximum linking density.  We find that varying the maximum linking density 
has little effect on the distribution of crossing numbers within voids.  We also 
examine the crossing number distribution as a function of distance from the void 
edge (depth) rather than distance from the void center.  While void depth has 
been used to provide a definition of a central region with a stronger preference 
for void-like galaxy properties \citep{Zaidouni24}, this did not improve the 
crossing number profiles of \Vsquared/VIDE voids.  We conclude that VoidFinder 
more effectively identifies the dynamically-distinct regions primarily occupied 
by dark matter particles with low crossing number.

The AbacusSummit Hugebase simulation analyzed in this work is one of the largest 
to be used in a crossing number analysis.  We modified the ORIGAMI algorithm 
(available for downloads at 
\url{https://github.com/dveyrat/origami/tree/subdivide}) to allow it to be run 
on this $2304^3$ particle set.  These modifications enable similar analyses 
using other AbacusSummit simulations, such as studies of the effects of 
different void-finding algorithms on cosmological constraints.

\section*{Acknowledgements}
The authors would like to thank Stephen W. O'Neill, Jr. for his help improving 
the VoidFinder algorithm in VAST, and Michael Vogeley and Fiona Hoyle for the 
original VoidFinder code and void-finding expertise.  D.V. and S.B. acknowledge 
support from the U.S. Department of Energy Office of High Energy Physics under 
the grant DE-SC0008475.  K.D. and D.V. acknowledge support from grant 62177 from 
the John Templeton Foundation.

This research used resources of the National Energy Research Scientific 
Computing Center (NERSC), a U.S. Department of Energy Office of Science User 
Facility located at Lawrence Berkeley National Laboratory.

\bibliographystyle{aasjournal}
\bibliography{Veyr0824}

\end{document}